\documentclass[preprint,12pt]{emulateapj}

\shorttitle{Relativisitic Velocities in 3C\,345}
\shortauthors{Roberts \& Wardle}
\usepackage{datetime}

\begin{document}

\title{Evidence for Highly Relativistic Velocities\\in the Kiloparsec-Scale Jet of the Quasar 3C\,345}

\author{David H.\ Roberts\footnote{Visiting Astronomer, National Radio Astronomy Observatory} \& John F.\ C.\ Wardle}
\affil{Department of Physics MS-057, Brandeis University, Waltham, MA 02454-0911}
\email{roberts@brandeis.edu}

\begin{abstract}

In this paper we use radio polarimetric observations of the jet of the nearby bright quasar 3C\,345 to estimate the fluid velocity on kiloparsec scales.
The jet is highly polarized, and surprisingly, the electric vector position angles in the jet are ``twisted'' with respect to the jet axis. Simple models of magnetized jets are investigated in order to study various possible origins of the electric vector distribution. In a cylindrically-symmetric transparent jet a helical magnetic field will appear either transverse or longitudinal due to partial cancellations of Stokes parameters between the front and back of the jet. Synchrotron opacity can break the symmetry, but it leads to fractional polarization less than that observed, and to strong frequency dependence that is not seen. Modeling shows that differential Doppler boosting in a diverging jet can break the symmetry, allowing a helical magnetic field to produce a twisted electric vector pattern. Constraints on the jet inclination, magnetic field properties, intrinsic opening angle, and fluid velocities are obtained, and show that highly relativistic speeds ($\beta \ga 0.95$) are required. This is consistent with the observed jet opening angle, with the absence of a counter-jet, with the polarization of the knots at the end of the jet, and with some inverse-Compton models for the X-ray emission from the 3C\,345 jet. This model can also apply on parsec scales and may help explain those sources where the electric vector position angles in the jet are neither parallel nor transverse to the jet axis. 

\end{abstract}


\section{INTRODUCTION}

An important question about extragalactic radio jets is whether or not the highly relativistic velocities often present on parsec scales continue to kiloparsec scales. Here we discuss a novel piece of information suggesting that this is indeed the case in at least one quasar jet.

3C\,345 is a superluminal quasar with inferred Lorentz factor $\Gamma$ up to at least 20 on parsec scales \citep{Lister}. In a companion paper \citep{RWM}, hereafter RWM, we present detailed VLA observations of the kiloparsec-scale jet of 3C\,345. Its most surprising feature is a ``twist'' to the electric vector distribution, suggesting a helical magnetic field. This was first seen by \citet{KWR}, and is unlike the jets in other Fanaroff-Riley Type~II radio sources, where the electric vectors are typically transverse to the jet axis, suggesting a longitudinal magnetic field \citep{BridlePerley}. In this paper we examine possible origins of this unusual electric vector distribution. Section~\ref{s:data} summarizes the relevant data on the 3C\,345 jet, \S\ref{s:models} describes simple models for the polarization of the jet, and \S\ref{s:concl} presents our conclusions.

\section{KEY DATA ON THE KILOPARSEC-SCALE JET OF 3C\,345}
\label{s:data}

\begin{figure}[t]{}
\includegraphics[width=1.0\columnwidth]{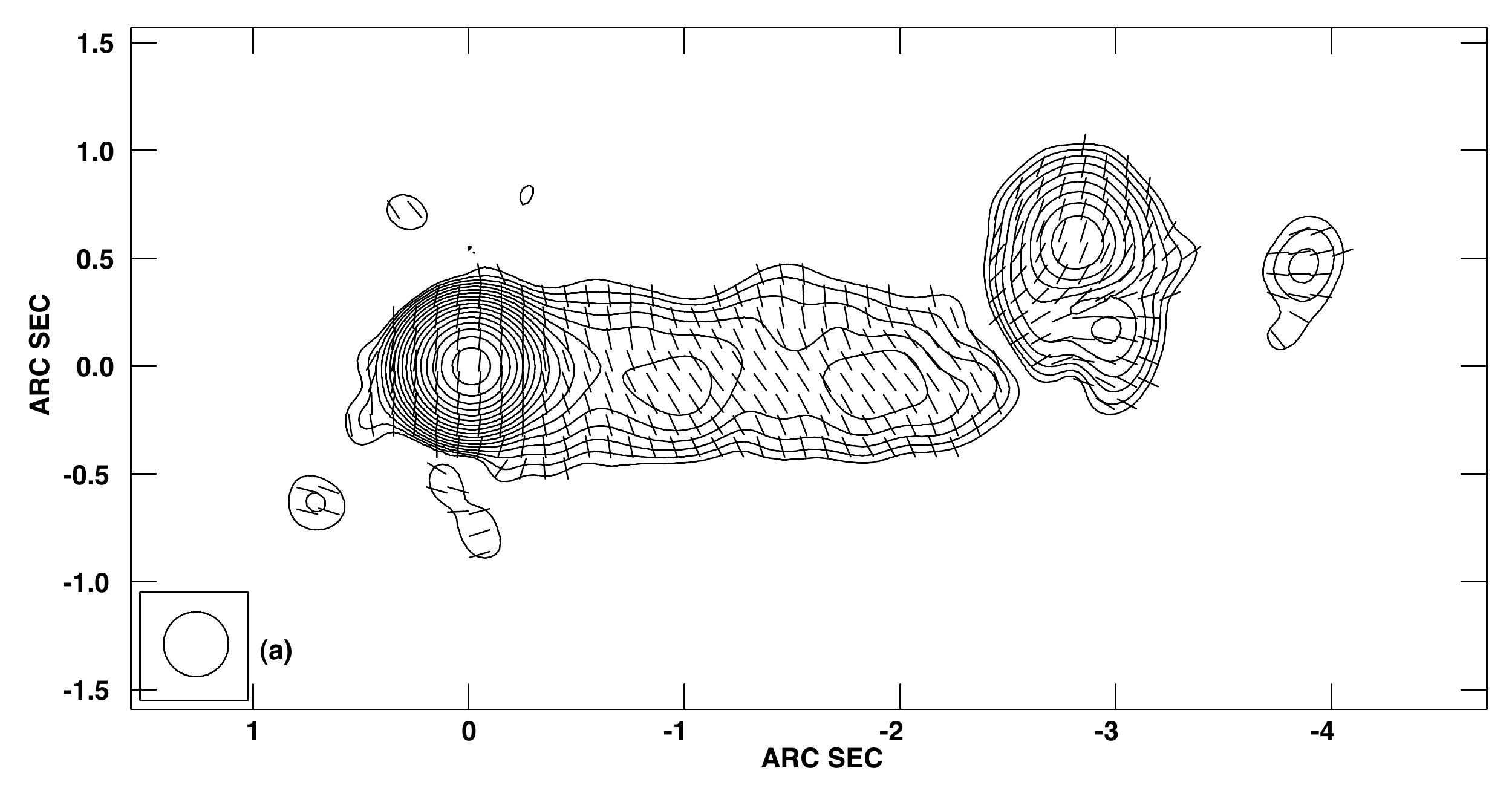}
\includegraphics[width=1.0\columnwidth]{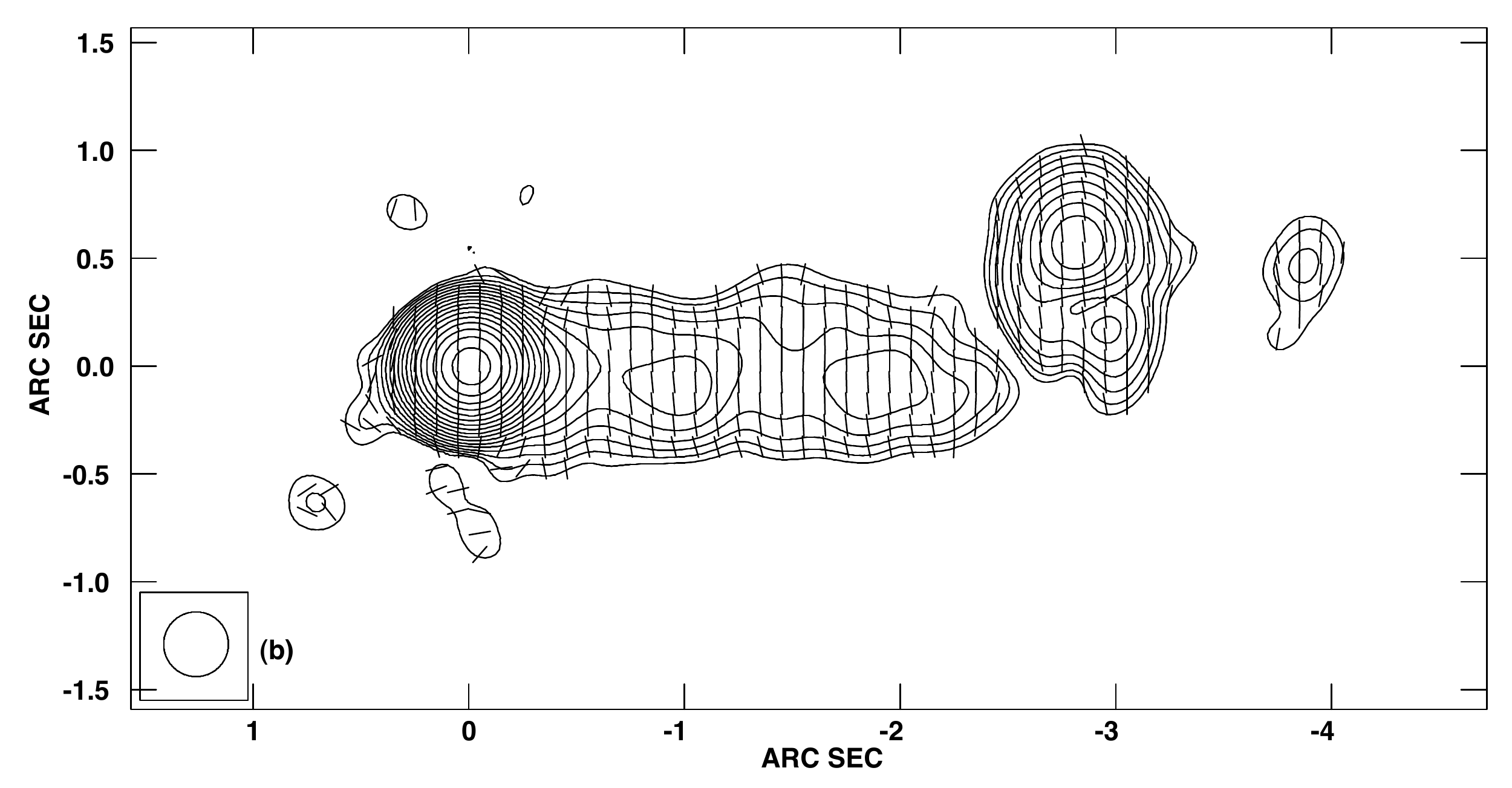}
\caption{Linear polarization of the kiloparsec-scale jet of 3C\,345 at 5~GHz. (a) Contours of linearly polarized intensity with ticks showing the orientation of the electric vectors. Note the unusual twist of the electric vectors in the main body of the jet. (b) Contours of linearly polarized intensity with ticks displaying the {\em differences} between the electric vector orientations at 5 and 8~GHz as position angle. Vertical lines indicate no Faraday rotation. These figures have been rotated by $-50^\circ$. From RWM.\label{fig:Poln}\\}
\end{figure}

Here we summarize the key features of the 3C\,345 jet as determined by RWM. (1) The fractional polarization in the jet is high, ranging from $0.2$ to $0.5$, and is systematically greater at the edges. (2) Unusually, the inferred magnetic field direction in the main body of the jet is neither longitudinal nor transverse, but makes an apparent helix (Figure~\ref{fig:Poln}a). At the center of the jet the twist is about $35^\circ$. (3) The ``twist'' in the apparent magnetic field is not due to Faraday rotation, which is a few degrees or less (Figure~\ref{fig:Poln}b). (4) The jet diverges slightly with an apparent semi-opening angle of about $\phi_a = 9.4^\circ$.  (5) The mean spectral index of the jet is 0.85 ($I_\nu \propto \nu^{-\alpha}$). (6) There is no counter-jet to a limit of 5\% of the brightness of the main jet.

\begin{figure}[t]{}
%\plottwo{542IQUshifted}{552IQUshifted}\\
%\plottwo{562IQUshifted}{572IQUshifted}\\
%\plottwo{582IQUshifted}{592IQUshifted}
\plottwo{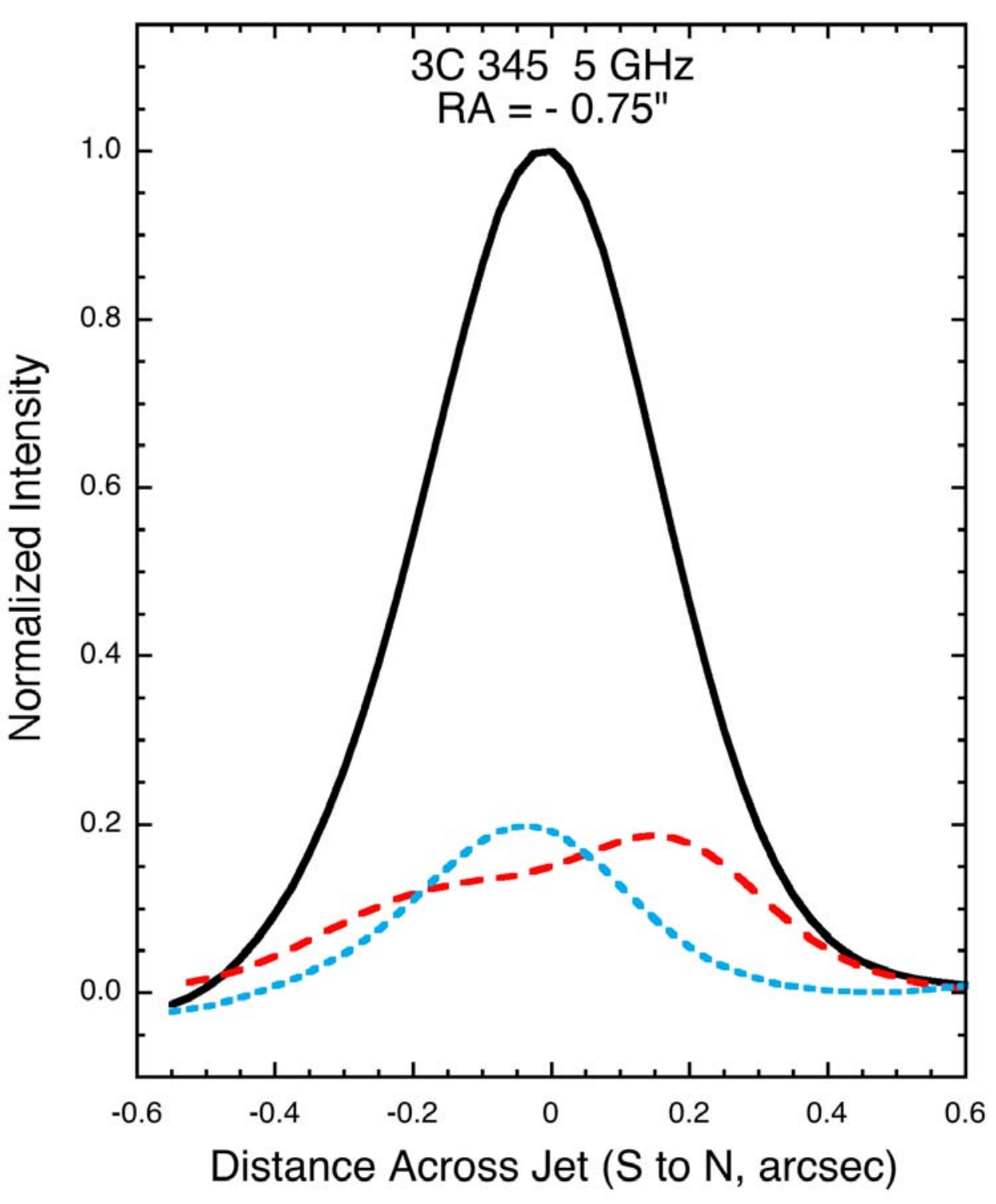}{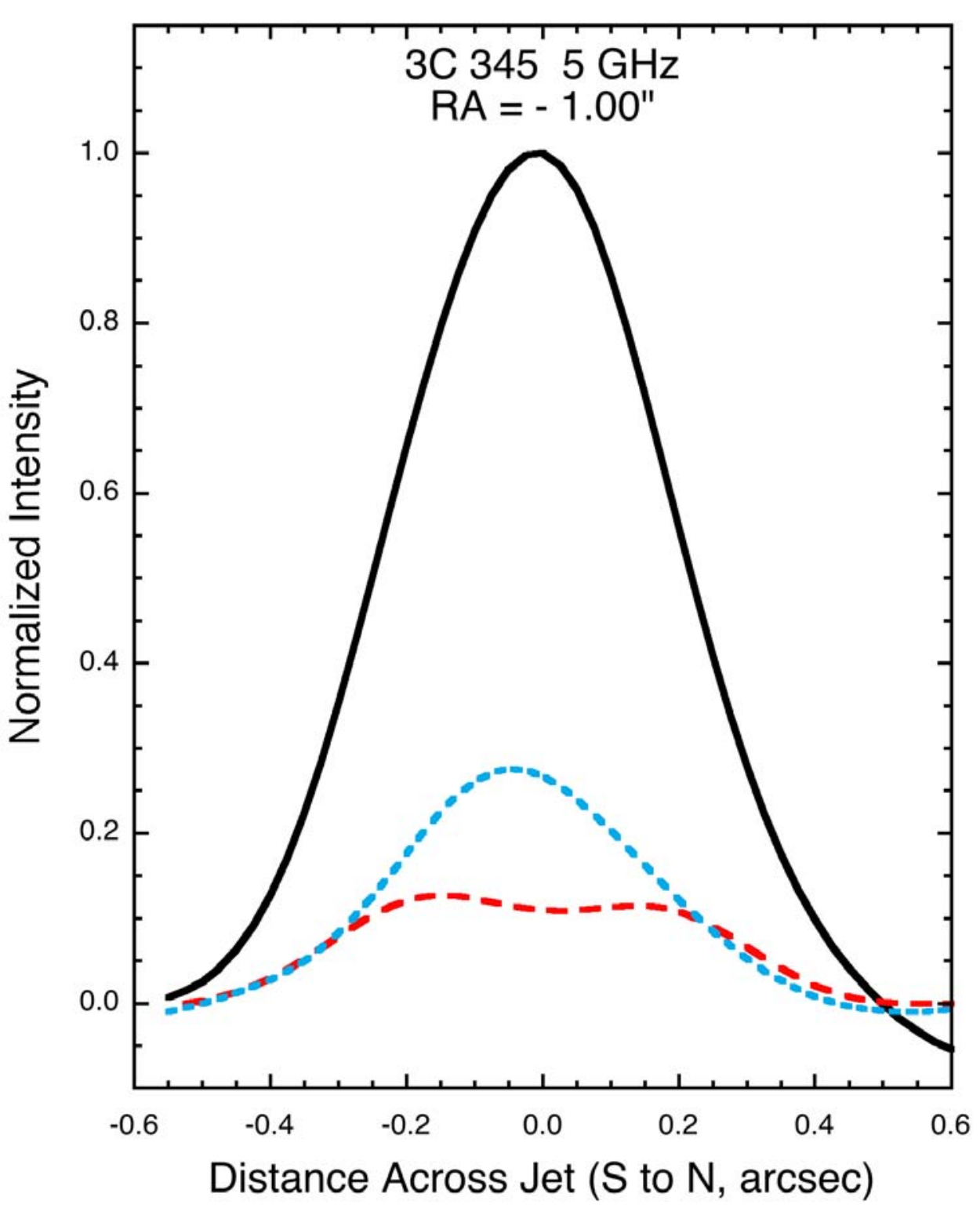}\\
\plottwo{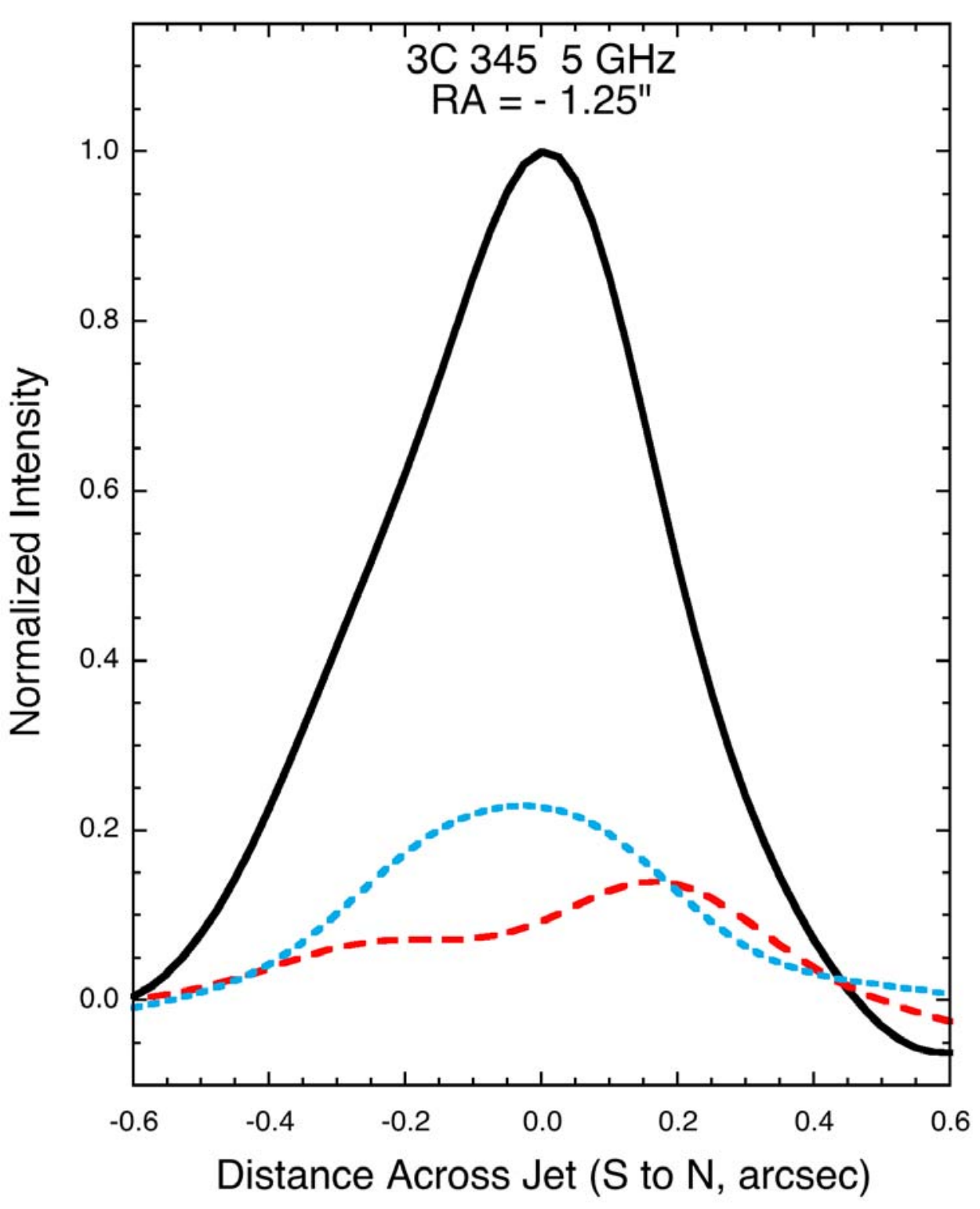}{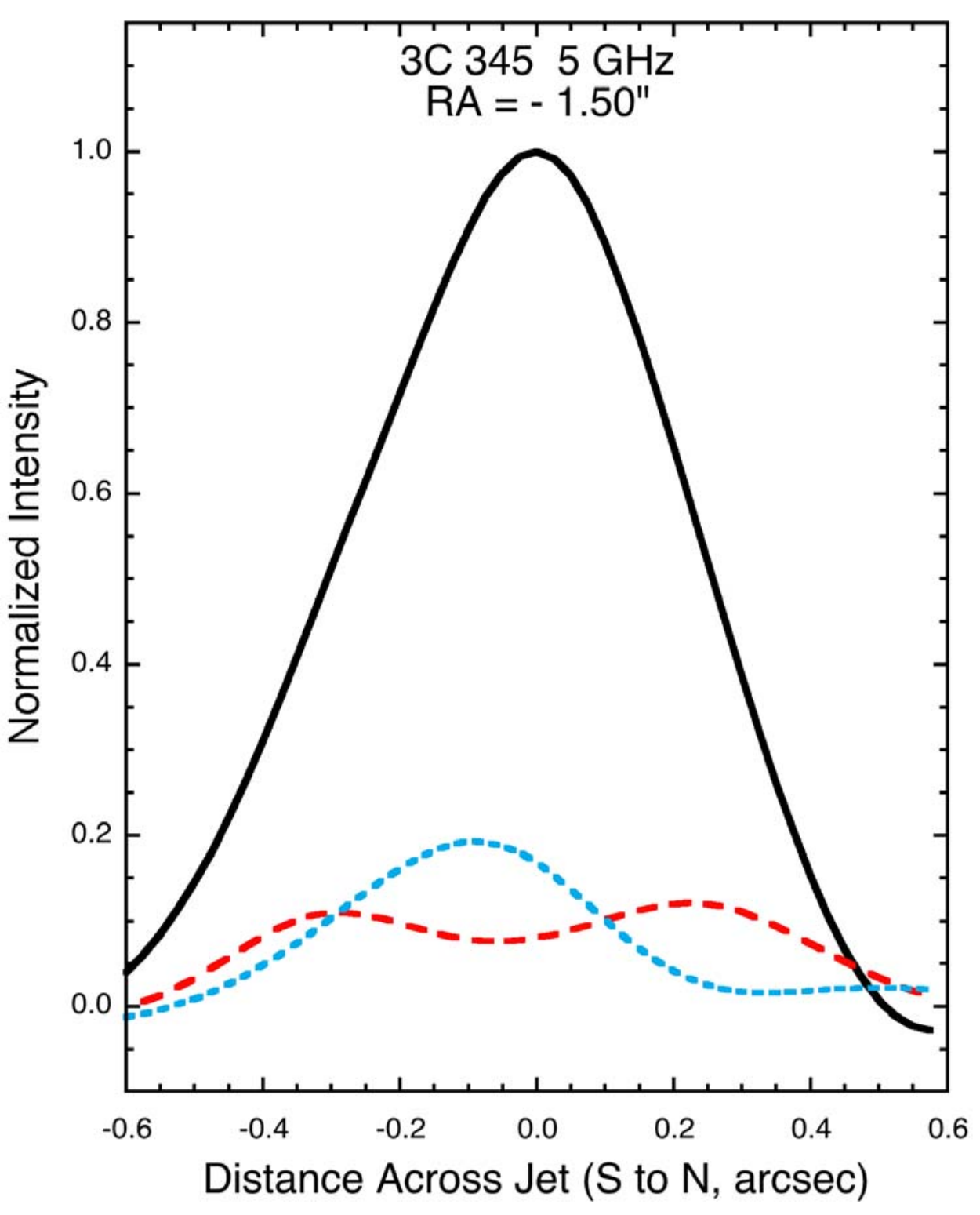}\\
\plottwo{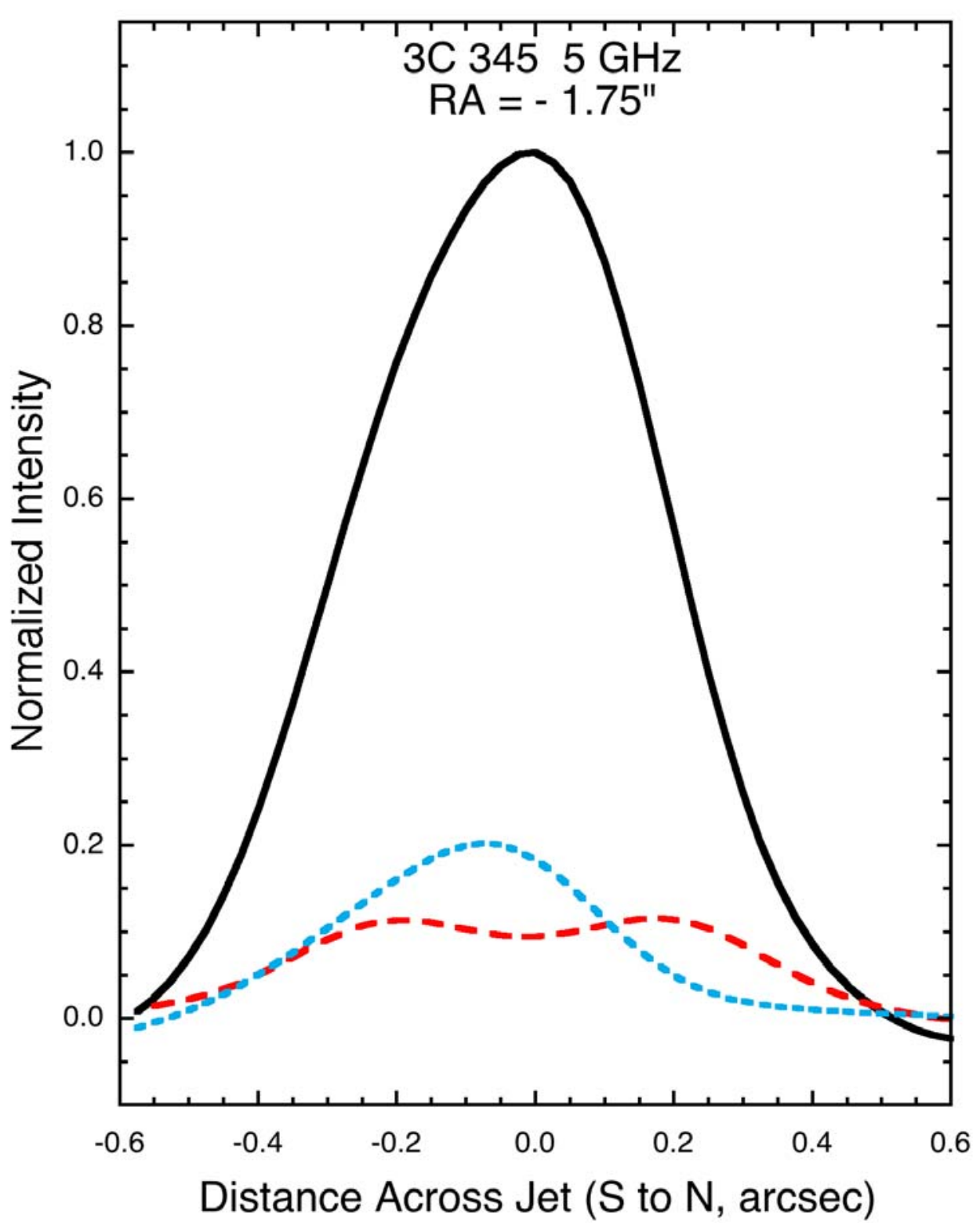}{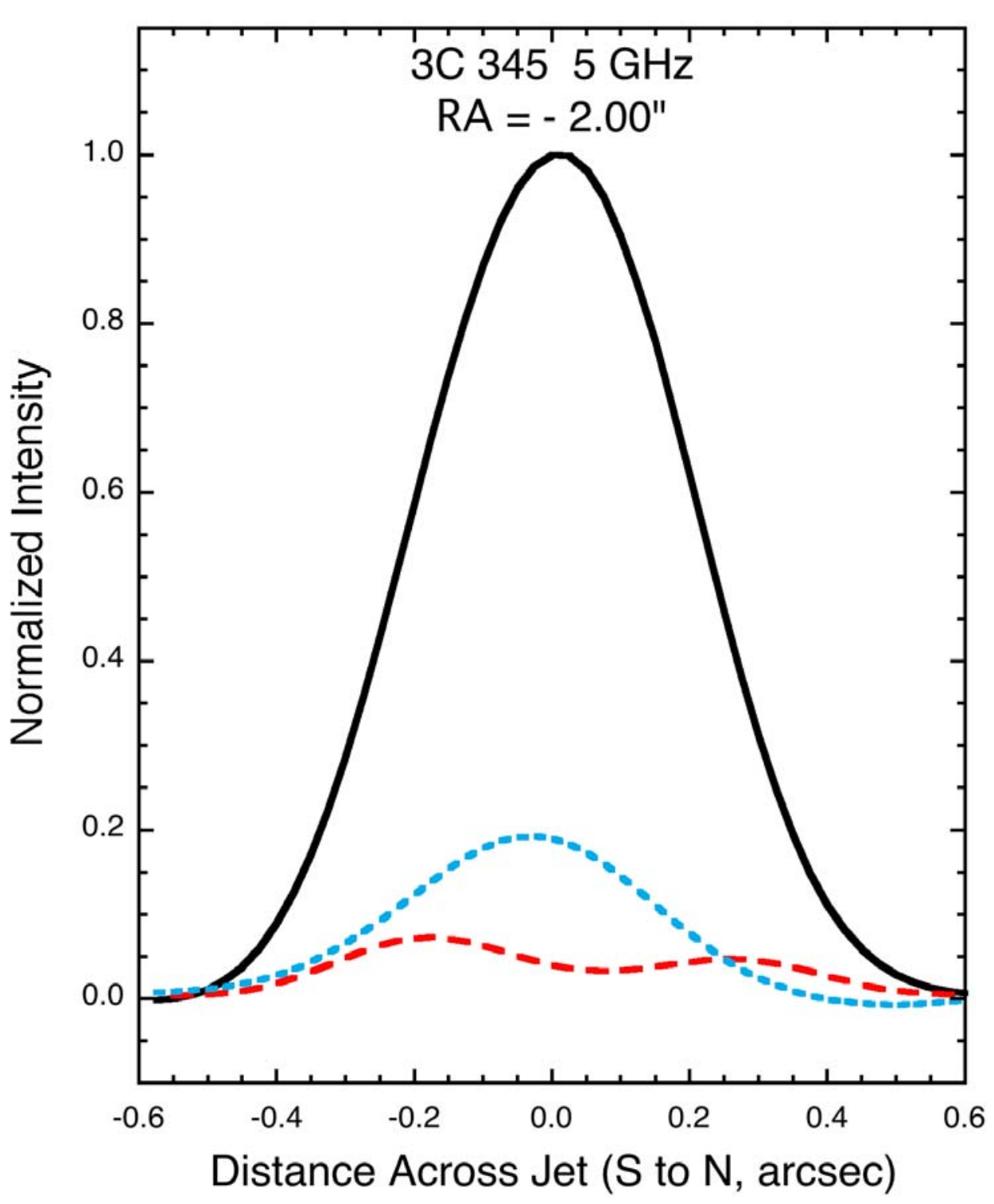}
\caption{Slices of $I/I_{peak}$ (black, solid), $Q/I_{peak}$ (red, broken), and $U/I_{peak}$ (blue, dotted) across the 3C\,345 jet at 5~GHz. The interval between slices, which range from $0.75\arcsec$ to $2.00\arcsec$ from the core, is $0.25\arcsec$. These data were derived from Figures~2 \& 9 in RWM. \\ \label{fig:slices}}
\end{figure}

In Figure~\ref{fig:slices} we show cross-sectional slices of the Stokes parameters $I$, $Q$, \& $U$ for six position in the jet. The purpose of this paper is to use these data to constrain fluid velocities on kiloparsec scales in 3C\,345.

\epsscale{1.0}

\epsscale{0.9}

\section{MODELS FOR THE TWISTED POLARIZATION\\OF THE 3C\,345 JET}
\label{s:models}

Is the magnetic field in the 3C\,345 helical, as it appears to be? There are few similar examples in the literature, the best observed one being M\,87. There the well-resolved jet shows a twisted filamentary structure in the radio \citep{Owen} that has been ascribed to a Kelvin-Helmholtz instability \citep{Hardee}, and the magnetic field structure seems to follow the filaments. Optically it looks quite similar to the radio image \citep{Fraix}. In 3C\,345 we see no filamentary structure in the jet, and suggest instead that the twisted electric field pattern is the result of relativistic effects. A truly helical field in a transparent homogeneous cylindrical jet would appear either transverse or longitudinal, not helical, due to cancellations between the back and the front (in a horizontal jet this means that $U$ is identically zero). Two possibilities suggest themselves; either the symmetry through the jet is broken by opacity or it is broken by relativistic effects. In the latter case, the gentle divergence in the profile of the jet demonstrated by RWM suggests a similarly diverging velocity field. In such a field, differential Doppler boosting due to differing line-of-sight velocity components in the front and back of the jet will break the symmetry, permitting there to be non-zero $U$, and thus creating a twist to the observed electric vectors if the underlying magnetic field is helical. Here we show that this is sufficient to produce the twist seen in 3C\,345, and use it to determine important physical parameters of the jet.

\subsection{Synchrotron Opacity}

We have made simple radiative transfer calculations using the equations in \citet{P70} that show that adding some synchrotron  opacity can produce an apparently-helical field, but that the fractional polarization is reduced well below the $\sim 25\%$ typical of the 3C\,345 jet, and that the polarization structure would be a strong function of frequency, which is not the case here. Thus we reject this explanation.

\subsection{Differential Doppler Boosting}

In order to assess the effects of differential Doppler boosting in a slightly conical jet, we have investigated the following model. The jet is taken to be a uniform-density  axisymmetric cylinder filled with relativistic electrons with energy index $1+2\alpha$ containing a helical magnetic field made up of two parts. The first is a toroidal component generated by a uniform current density, and the second component is a uniform longitudinal field; we parameterize the helicity of the field by the ratio $b$ of longitudinal component to the toroidal component at the surface. The effect of Doppler boosting is included in an {\em ad hoc} manner by superimposing a diverging velocity field whose outer profile is constrained by the observed opening angle of the jet. The angle of the velocity vectors to the jet axis varies with normalized radius $r$ as
$$
\eta(r) = \phi_i r^\epsilon
$$
where the radial parameter satisfies $0 \leq r \leq 1$, $\phi_i$ is the intrinsic half-opening angle of the jet, and $\epsilon$ is an adjustable parameter that we take to be unity. The radiative transfer was done in the observer frame using the equations in \cite{P70}, incorporating the relationship between the source magnetic field and radiation electric vectors in a relativistically-moving medium \citep{LPB}. 

If the fluid speed is $\beta c$ and the inclination of the jet axis to the line of sight is $i$, in the fluid frame the photon paths make an angle $\theta^\prime$ with the jet axis, where
$$
\cos{\theta^\prime} = \frac{\cos{i}-\beta}{1-\beta \cos{i}}.
$$
The computations show that the symmetry of the profiles of $I$, $Q$, and $U$ across the jet constrain the angle $\theta^\prime$ to be very close to $90^\circ$. This means that $\sin i \simeq 1/\Gamma$, so $\beta$ and $i$ cannot be chosen independently. The intrinsic opening angle of the jet $\phi_i$ and the apparent  opening angle $\phi_a$ are related by geometry according to
$$
\tan \phi_i = \tan \phi_a \sin i,
$$
which means that increasing the speed of a model jet reduces the intrinsic opening angle of the jet for a given observed opening angle, partially counteracting the increased differential Doppler effect. One upshot of this is that the maximum $U/I$ that can be generated is sensitive to $\beta$ only linearly, and is not explicitly a function of $\Gamma$.

%The model is constrained by the following considerations. (1) The observed  jet opening angle must be consistent with that measured in RWM. (2) The fractional polarization must be about 25\% in the center of the jet and be systematically higher toward the edges. (3) The electric vector angle ``twist'' must be about $35^\circ$ in the center of the jet, and fall to zero at the edges. (4) The model must predict a jet--counter-jet ratio consistent with the conclusions of RWM. 

\begin{figure}[t]{}
\plotone{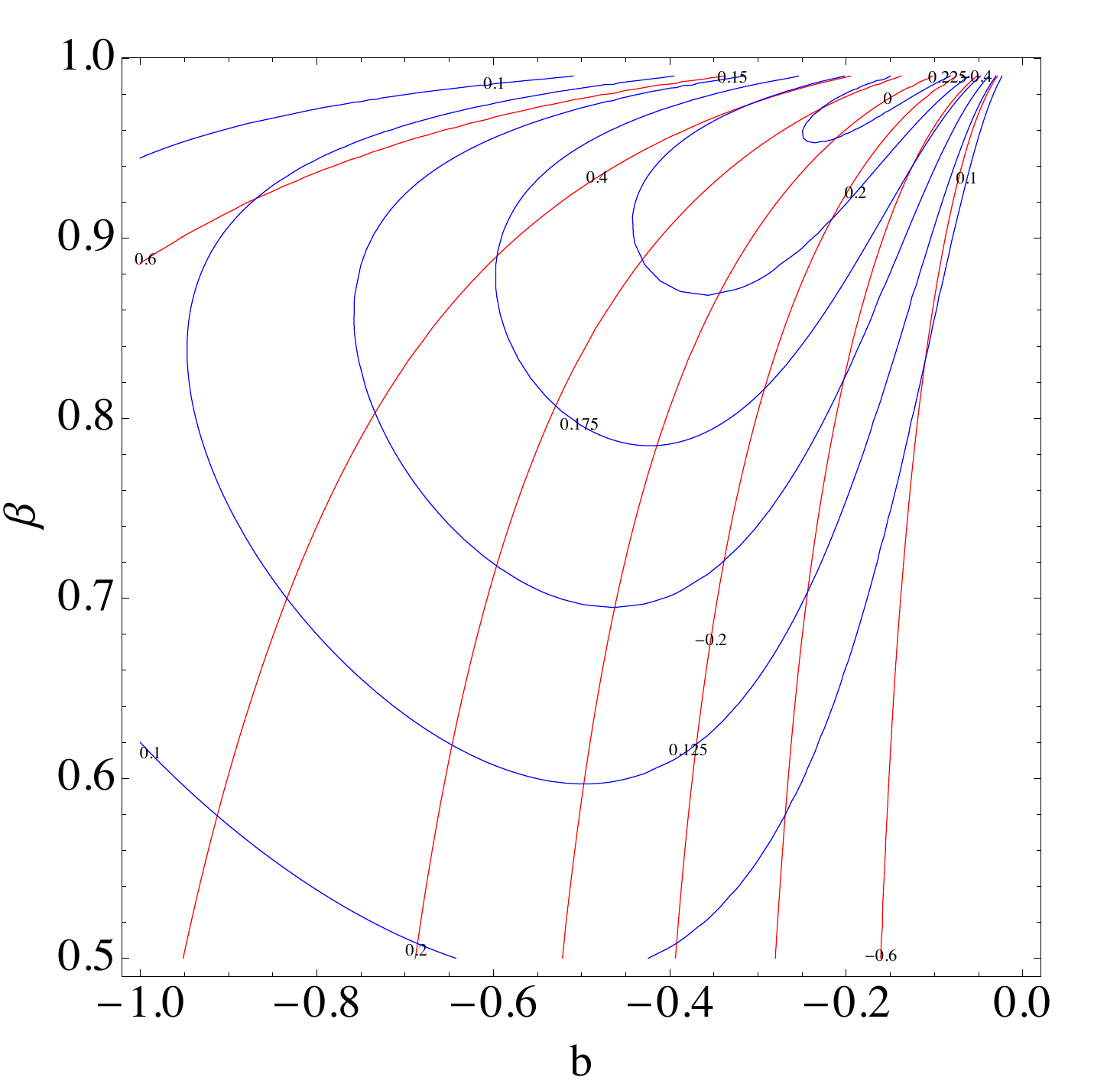}
%\plotone{Contours}
%\plotone{Contours2}
%\plotone{ContourDetails}
\caption{Model curves of constant $Q/I$ (red,) and $U/I$ (blue) at the center of the jet, as functions of the parameters $b$ and $\beta$, assuming $\phi_a = 9.4^\circ$.\label{fig:ContourPlot}}
\end{figure}

\begin{figure}[t]{}
\plotone{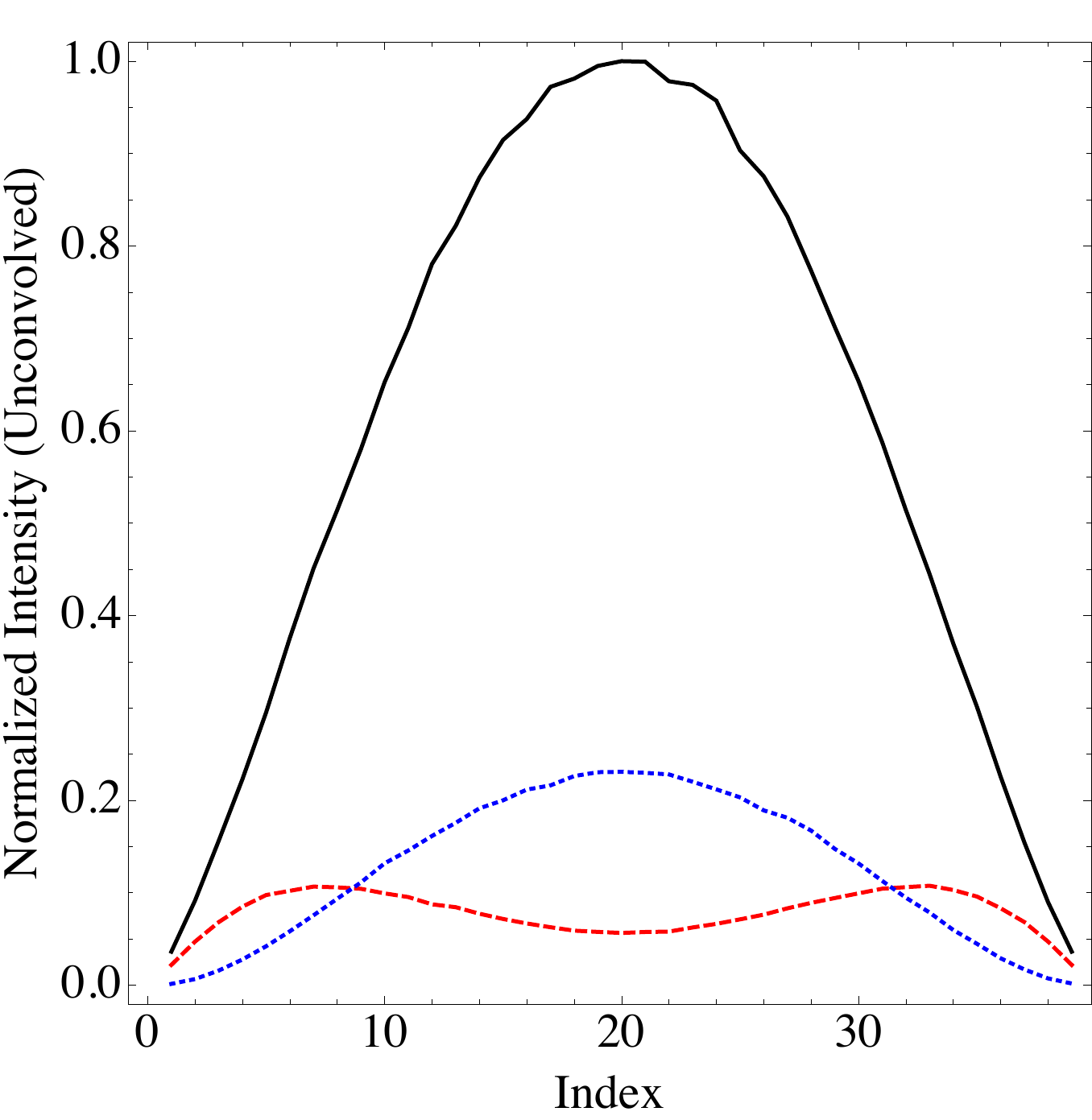}
%\plotone{BestModelUnconvolved}
%\plotone{UnconvolvedModel}
\caption{Nominal model for the Stokes parameters in 3C\,345, without convolution with the observing beam. Here $I/I_{peak}$ is black, $Q/I_{peak}$ is red (broken), and $U/I_{peak}$ is blue (dotted).\label{fig:ModelUnconv}}
\end{figure}

\begin{figure}[t]{}
\plotone{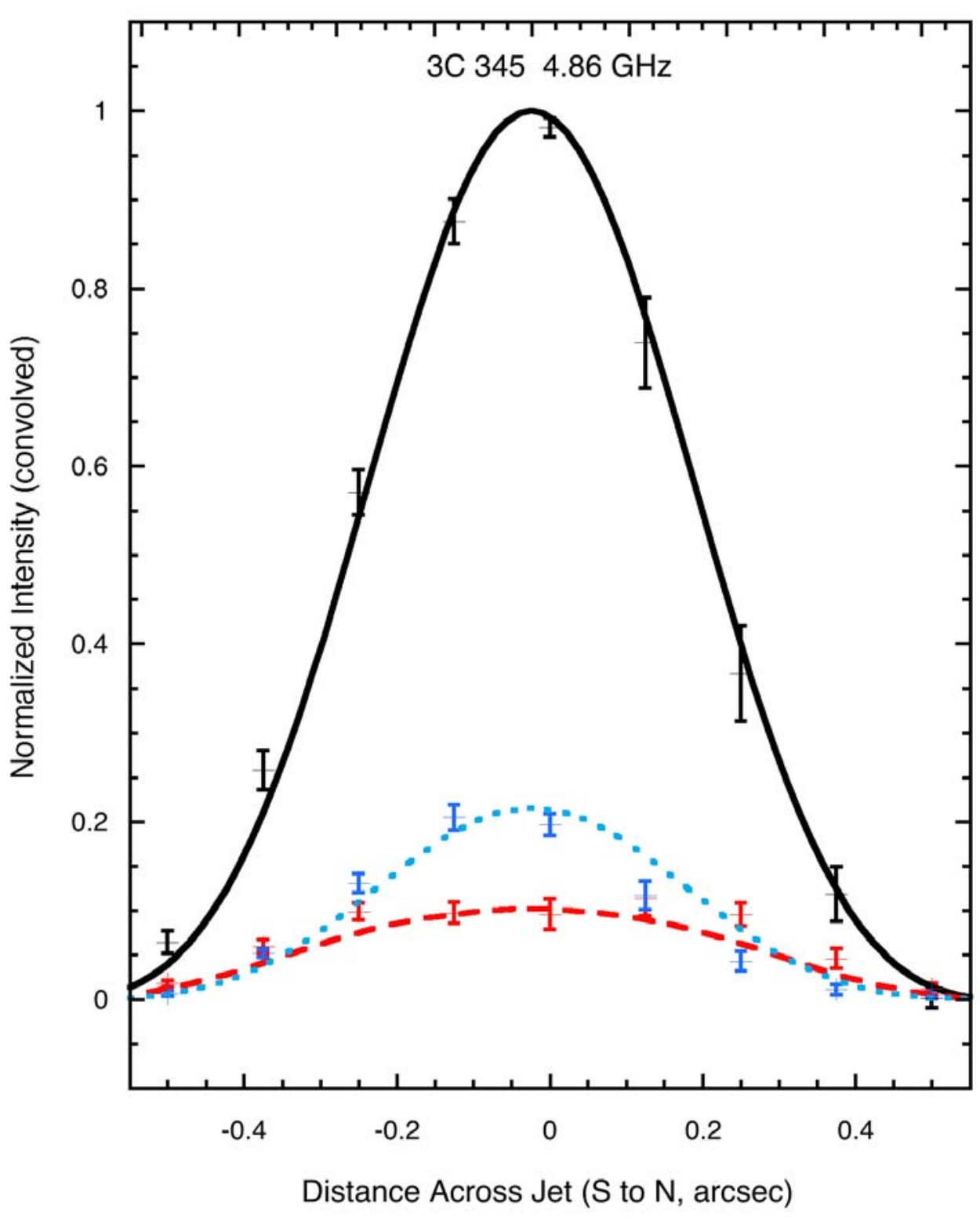}
%\plotone{BestModel}
%\plotone{Data+W1Model}
\caption{Nominal model for the linearly polarized Stokes parameters of the arcsecond jet of 3C\,345 convolved with the observing beam plotted against the data at 5~GHz. The data are the means and standard errors of the four slices between $1.0\arcsec$  and $1.75\arcsec$ from the core. Here $I/I_{peak}$ is black, $Q/I_{peak}$ is red (broken), and $U/I_{peak}$ is blue (dotted). \\ \label{fig:models}}
\end{figure}

We searched for solutions in the $b$--$\beta$ plane using contours of constant $Q/I$ and $U/I$ at the center of the jet, where the values should be approximately $0.05$ and $0.22$, respectively, prior to convolution with the beam (see Figure~\ref{fig:ContourPlot}). The nominal model without convolution with the observing beam is shown in Figure~\ref{fig:ModelUnconv}; its parameters are $\beta=0.97$, $i=14^\circ$, $\phi_i = 2.3^\circ$, and $b = -0.19$. Figure~\ref{fig:models} shows the average observed profiles of $I$, $Q$, and $U$, each normalized by the peak of $I$, as functions of distance across the jet versus the predictions of this model when it is convolved with the observing beam of RWM. We find the following. (1) It is possible to choose parameters $\beta$ and $b$ for the jet that produce Stokes parameters that closely match those of the 3C\,345 jet. (2) The ratio of surface longitudinal to toroidal magnetic field strength in the observer frame must satisfy $-0.3 \lesssim b \lesssim -0.1$. (3) The inclination of the jet to the line of sight is $8^\circ \lesssim i \lesssim 16^\circ$. (4) The fluid speed is $\beta \ga 0.95$.

\epsscale{1.0}

\epsscale{1.0}

\epsscale{1.0}

We also did numerical and analytic calculations in the plasma frame that were consistent with these results when transformed to the observer frame (see \citet{CH}). For example, in the special case of a spectral index of $\alpha = 1$ and $\theta^\prime = 90^\circ$, it can be shown that the normalized profiles of $I$, $Q$, and $U$ are
$$
\frac{I(x)}{I(0)} \simeq \frac{3b^{\prime 2}+1-x^2}{1+3b^{\prime 2}} (1-x^2)^{1/2} + O((\phi_a \beta)^2),
$$
$$
\frac{Q(x)}{I(0)} \simeq \frac{3}{4}  \frac{3b^{\prime 2}-1+x^2}{1+3b^{\prime 2}} (1-x^2)^{1/2}  + O((\phi_a \beta)^2),
$$
$$
\frac{U(x)}{I(0)} \simeq - \frac{9 b^\prime \beta \phi_a}{2(1+3b^{\prime 2})} (1-x^2)^{3/2} + O((\phi_a \beta)^3).
$$
Here $-1 \le x \le 1$ is normalized position across the jet and $b^\prime =  \Gamma b $ is the pitch angle parameter in the frame of the fluid. These profiles agree very well with numerical integrations for the same parameters.

For such a simple model the success is gratifying. All of the essential features of the linear polarization of the jet are reproduced within the observational uncertainties. Models without highly relativisitic bulk flows are unable to reproduce the twist of the electric vectors, and we regard the fit of data and model to be strong evidence for such flows on kiloparsec scales in 3C\,345. All of this assumes that the apparent semi-opening angle of the jet is about $9^\circ$ (RWM). Since $U/I$ at the center of the jet is proportional to $b \beta \phi_a$ and $b$ is constrained by the observed values of $Q/I$, even a 20\% reduction in $\phi_a$ renders the models untenable.

If 3C\,345 is typical, and helical magnetic fields are present is all kiloparsec-scale jets, we must explain why we don't see twists in other highly-polarized FR~II jets. One possibility is that 3C\,345 is core-dominated and thus is inclined near the line of sight, while those observed by, e.g., \citet{Bridle94}, are lobe-dominated and their jets are inclined at much larger angles to the line of sight. When we apply our model for 3C\,345 to such jets the result is apparently longitudinal magnetic fields.

The symmetry of $I$, $Q$, \& $U$ across the jet means that it is seen at an angle $\theta^\prime \simeq 90^\circ$ in the frame of the fluid, and that the inclination to the line of sight is $i \simeq 1/\Gamma$. Since this is the condition for maximal superluminal motion, it suggests that deep polarimetric observations of the kiloparsec-scale jets in other core-dominated and/or superluminal sources be undertaken to search for further examples of twisted electric vector configurations.

\section{CONCLUSIONS}
\label{s:concl}

Possible origins of the twisted electric vector orientations in the kiloparsec-scale jet of 3C\,345 are investigated, and it is found that differential Doppler boosting in a conical jet is adequate to permit a helical magnetic field to produce a twisted electric field pattern. Constraints on the jet speed and inclination to the line of sight are derived, and are found to be consistent with the presence of superluminal motion on parsec scales and with the absence of a counter-jet on kiloparsec scales. Acceptable models have speed $\beta \ga 0.95$, intrinsic jet opening angle $\phi_i \lesssim 2.6^\circ$, and inclination to the line of sight $i \lesssim 16^\circ$. Models with $\beta \lesssim 0.95$ are unsuccessful, so we conclude that the fluid speeds in the kiloparsec-scale jet of 3C\,345 are highly relativistic. This is in agreement with the conclusions drawn from inverse-Compton models of the X-ray emission seen from 3C\,345 by Chandra \citep{CHANDRA}. This model can also apply on parsec scales, and may help explain those sources where the electric vector position angles in the parsec scale jet are neither parallel nor transverse to the direction of the jet (\cite{CWRG}; \citet{GPC}; \citet{PTZ}; \citet{LH}).

\section{ACKNOWLEDGMENTS}

The National Radio Astronomy Observatory is a facility of the National Science Foundation operated under cooperative agreement by Associated Universities, Inc. D.~H.~R.\ gratefully acknowledges the support of the William R.\ Kenan, Jr.\ Charitable Trust and of the National Radio Astronomy Observatory.  J.~F.~C.~W.\ is supported by NSF grant AST-1009262. Any opinions, findings, and conclusions or recommendations expressed in this material are those of the authors and do not necessarily reflect  the views of the National Science Foundation. We thank Jean Eilek for helpful conversations and a careful reading of the manuscript.


\begin{thebibliography}{} % In [], must be NO SPACE between last character of name and the "(" of date!
\bibitem[Bridle et al.(1994)]{Bridle94} Bridle, A.\ H., Hough, D.\ H., Lonsdale, C.\ L., Burns, J.\ O., \& Laing, R.\ A. 1994, \aj, 108, 766
\bibitem[Bridle \& Perley(1984)]{BridlePerley} Bridle, A.\ H., \& Perley, R.\ A. 1984, \araa, 22,319
\bibitem[Cawthorne et al.(1993)]{CWRG} Cawthorne, T.\ V., Wardle, J.\ F.\ C., Roberts, D.\ H. \& Gabuzda, D.\ C.\ 1993, \apj, 416, 519
\bibitem[Cocke \& Holm(1972)]{CH} Cocke, W.\ J., \& and Holm, D.\ A.\ 1972, Nature, 240, 161
\bibitem[Fraix-Burnet, Le Borgne, \& Nieto(1989)]{Fraix} Fraix-Burnet, D., Le Borgne, J.-F. \& Nieto, J.-L. 1989, A\&A, 224, 17.
\bibitem[Hardee \& Eilek(2011)]{Hardee} Hardee, P.\ E. \& Eilek, J.\ A. 2011, \apj, 735, 61
\bibitem[Gabuzda, Pushkarev, \& Cawthorne(2000)]{GPC} Gabuzda, D.\ C., Pushkarev, A.\ B. and Cawthorne, T.\ V.\ 2000, \mnras, 319, 1109
\bibitem[Kharb et al.(2012)]{CHANDRA} Kharb, P., Lister, M.~L., Marshall, H.~L., \& Hogan, B.~S. 2012, \apj, 748, 81
\bibitem[Kollgaard, Wardle, \& Roberts(1989)]{KWR} Kollgaard, R.\ I., Wardle, J.\ F.\ C., \& Roberts, D.\ H. 1989, \aj, 97, 1550
\bibitem[Laing(1980)]{L80} Laing, R. A. 1980, \mnras, 193, 439
\bibitem[Lister \& Homan(2005)]{LH} Lister, M.\ L., \& Homan, D.\ C.\ 2005, \aj, 130, 1389
\bibitem[Lister et al.(2009)]{Lister} Lister, M.\ L. et al. 2009, \aj, 138, 1874
%\bibitem[Laing(1981)]{L81} Laing, R. A. 1981, \apj, 248, 87
%\bibitem[Laing, Canvin, \& Bridle(2006)]{LCB} Laing, R. A., Canvin, J. R., \& Bridle, A. H. 2006, Astron. Nach., 327, 523
\bibitem[Lyutikov, Pariev, \& Blandford(2003)]{LPB} Lyutikov, M., Pariev, V.~I., \& Blandford, R.~D. 2003, \apj, 597, 998
\bibitem[Owen, Hardee, \& Cornwell(1989)]{Owen} Owen, F.\ N., Hardee, P.\ E. \& Cornwell, T.\ J. 1989, \apj, 340, 698
\bibitem[Pacholczyk(1970)]{P70} Pacholczyk, A.\ G.\ 1970, {\em Radio Astrophysics}, Freeman (San Francisco)
\bibitem[Pollack, Taylor, \& Zavala(2003)]{PTZ} Pollack, L.\ K., Taylor, G.\ B., \& Zavala, R.\ T.\ 2003, \apj, 589, 733
\bibitem[Roberts, Wardle, \& Marchenko(2012)]{RWM} Roberts, D.~H., Wardle, J.\ F.\ C., \& Marchenko, V. V.\  2012, \aj, submitted (arXiv:???) (RWM)
\end{thebibliography}
\end{document}